\documentclass[letterpaper, 10pt, conference]{IEEEtran}
\ifCLASSINFOpdf
\usepackage[pdftex]{graphicx}
\usepackage{subcaption}
\graphicspath{{images/}}

\DeclareGraphicsExtensions{.pdf,.jpeg,.png}
\else
\usepackage[dvips]{graphicx}
\graphicspath{eps}
\DeclareGraphicsExtensions{.eps}
\fi

\usepackage{stfloats}
\usepackage{float}
\usepackage{enumerate}
\usepackage{setspace}
\usepackage{balance}
\usepackage{caption}
\usepackage{xcolor}
\usepackage{graphicx}
\usepackage{algpseudocode}
\usepackage{color}
\usepackage{multicol}
\usepackage{multirow}
\usepackage[linesnumbered,ruled,vlined]{algorithm2e}
\usepackage[flushleft]{threeparttable}

\usepackage{setspace}
\usepackage{balance}
\usepackage[utf8]{inputenc}

\usepackage[scaled]{beramono}
\usepackage[T1]{fontenc}
\usepackage{setspace}
\usepackage{cite}
\usepackage{amsmath}
\usepackage{amsthm}
\usepackage{listings}
\usepackage{textcomp}
\usepackage[utf8]{inputenc}
\usepackage{tabulary}
\usepackage{enumitem}
\usepackage[hyphens]{url}
\usepackage[hidelinks]{hyperref}

\SetKwInput{KwInput}{Input}
\SetKwInput{KwOutput}{Output}
\DeclareUnicodeCharacter{00A0}{ }

\usepackage{draftwatermark}
\SetWatermarkScale{1}
\SetWatermarkColor[gray]{0.9}

\usepackage{tikz}
\newcommand*\circled[1]{\tikz[baseline=(char.base)]{
		\node[shape=circle,draw,inner sep=1pt] (char) {#1};}}

\def\baselinestretch{0.98}

\newif\ifconf
\conftrue

\begin{document}
	\title{BARISTA: Efficient and Scalable Serverless Serving System for Deep Learning Prediction Services} 	

	
\author{
	\IEEEauthorblockN{Anirban Bhattacharjee, Ajay Dev Chhokra, Zhuangwei Kang, Hongyang Sun, Aniruddha Gokhale, Gabor Karsai}
	\IEEEauthorblockA{Department of Electrical Engineering and Computer Science,
		Vanderbilt University, Nashville, Tennessee, USA\\
		Email: {\{anirban.bhattacharjee; ajay.d.chhokra; zhuangwei.kang; hongyang.sun; a.gokhale; gabor.karsai\}}@vanderbilt.edu
}}

\maketitle
\begin{abstract}
Pre-trained deep learning models are increasingly being used to offer a variety of compute-intensive predictive analytics services such as fitness tracking, speech and image recognition. The stateless and highly parallelizable nature of deep learning models makes them well-suited for serverless computing paradigm. However, making effective resource management decisions for these services is a hard problem due to the dynamic workloads and diverse set of available resource configurations that have various deployment and management costs. To address these challenges, we present a distributed and scalable deep-learning prediction serving system called Barista and make the following contributions. First, we present a fast and effective methodology for forecasting workloads by identifying various trends.  Second, we formulate an optimization problem to minimize the total cost incurred while ensuring bounded prediction latency with reasonable accuracy.  Third, we propose an efficient heuristic to identify suitable compute resource configurations. Fourth, we propose an intelligent agent to allocate and manage the compute resources by horizontal and vertical scaling to maintain the required prediction latency. Finally, using representative real-world workloads for an urban transportation service, we demonstrate and validate the capabilities of Barista.
\end{abstract}

\renewcommand\IEEEkeywordsname{Keywords}
	

\section{Introduction}
\label{sec:introduction}

\subsection{Emerging Trends}
Cloud-hosted, predictive analytics services based on \emph{pre-trained deep learning models}~\cite{lecun2015deep} have given rise to a diverse set of applications, such as speech recognition, natural language processing, fitness tracking, online recommendation systems, fraudulent behavior detection, genomics, and computer vision.  End-users of these services query these pre-trained models using an interactive web or mobile interface through RESTful APIs. Based on the supplied input, these pre-trained models infer the target values and return the prediction results to the end-users.   As an example, a speech recognition system transcribes spoken words into text~\cite{bahdanau2016end}.

These prediction services are usually containarized~\cite{al2017autonomic} and encapsulated with all the required software packages. Thus, deployer of these services can preferably use the \emph{function-as-a-service} (FaaS) approach to hosting these services in an event-driven manner, wherein the functions are executed on the occurrence of some trigger or event (e.g., incoming request).  This overall approach can be handled using \emph{serverless computing}~\cite{baldini2017serverless}, since the service creator needs only to provide the function logic, the trigger conditions, and the \emph{service level objectives} (SLOs), such as latency bounds, which are on order of few seconds. It is then the responsibility of the serverless platform provider to provide the hosting environment for these services and to ensure that the SLOs are met.

\subsection{ Challenges and State-of-the-Art}
The execution environments for deep learning-based prediction services typically comprise containers running on a cluster of virtual machines (VMs). These prediction services are usually stateless, parallelizable (multi-threaded) and compute-intensive. The model sizes of these prediction services are large (hundreds of megabytes to gigabytes), which takes a significant time to load them into the containers and provision the infrastructure.  Moreover, due to the parallelizable nature of these models, their running times can be substantially reduced by assigning more CPU cores (see Figure~\ref{Fig:modeltime}). However, allocating more memory only marginally improves the running times. Thus, a na\"{\i}ve approach to assuring the SLOs is to over-provision the service infrastructure; however, doing so imposes undue costs on the serverless provider.  Efficient management of computing resources dynamically is required to minimize the cost of hosting these services in the cloud ~\cite{gujarati2017swayam,baldini2017serverless,crankshaw2017clipper}.

Although a substantial amount of literature exists on finding the sweet spot between resource over-provisioning (which wastes resources and increases the cost) and under-provisioning (which violates the SLOs)~\cite{alipourfard2017cherrypick,delimitrou2014quasar,venkataraman2016ernest},
these works focus primarily on long-running analytics jobs for which the goal is to find optimal configurations to meet the SLOs and scale the resources dynamically to handle their variable workloads.

In contrast, the prediction services have short running times. Moreover, the incoming request (workloads) patterns can fluctuate significantly and follow a diurnal model, which requires rapid management of resources to meet the variable workload demands.  A reactive approach is not suitable as the prediction latency may increase significantly due to infrastructure provisioning time (e.g., order of minutes due to VM creation and model loading times).  Hence, the desired solution is one that can forecast the workload patterns and can estimate the required resources for the application to maintain the SLOs under the forecasted workload.   Determining the right cluster configurations and allocating the resources dynamically is hard due to the variability of cloud configurations (VM instance types), the number of VMs, and their deployment and management costs~\cite{alipourfard2017cherrypick,barker2010empirical}.

Only recently have some solutions started to emerge to address these concerns~\cite{gujarati2017swayam,ukidave2016mystic}.  Nevertheless, there remain many unresolved problems.  First, current horizontal elasticity solutions for prediction services often do not account for the container-based service lifecycle states (e.g., whether the VM is already up or not, or whether the container is running, and if so whether the model is loaded or not).  Each such state incurs a hosting cost and impacts the running time. Second, vertical elasticity solutions for containers do not yet fully exploit the parallelizable aspects of the pre-trained models.  Third, proactive resource scaling decisions require effective workload forecasting and must be able to monitor the service lifecycle states, both of which are missing in prior efforts.  Finally, existing strategies for container allocation tend to overlook performance interference issues from other co-located containerized services, which may cause unpredictable performance degradation.


\subsection {Solution Approach }
To address these unresolved problems, we present a serverless framework, called Barista, which hosts containerized, pre-trained deep learning models for predictive analytics in the cloud environment, and meets the SLOs of these services while minimizing hosting costs. Barista comprises an efficient, data-driven and scalable resource allocator, which estimates the resource requirements ahead of time by exploiting the variable patterns of the incoming requests, and a forecast-aware scheduling mechanism, which improves resource utilization while preventing physical resource exhaustion.  We show how serverless computing concepts for dynamically scaling the resources vertically and horizontally can be utilized under different scenarios.  Barista efficiently and cost-effectively provisions (scale-up and scale-down) resources for a prediction analytics service to meet its prediction latency bound.  Specifically, we make the following contributions:

\begin{enumerate}
\item \emph{\textbf{Workload Forecast:}} We propose a hierarchical methodology to forecast the workload based on historical data.
	
\item \emph{\textbf{Resource Estimation:}} Barista allows service providers to communicate the performance constraints of their service models regarding their SLOs.  An analytical model is provided to predict resource estimation based on latency bound, workload forecasting and the profiled execution time model on different cloud configurations.
	
\item \emph{\textbf{Serverless Resource Allocation:}} Barista provides a novel mechanism using serverless paradigm to allocate resources proactively based on the difference between resource requirement estimation and current infrastructure state in an event-driven fashion.
\end{enumerate}

\subsection {Organization of the Paper}
The rest of the paper is organized as follows:
Section~\ref{sec:survey} presents a survey of existing solutions in the literature and compares them with Barista;
Section~\ref{sec:problem} presents the problem formulation;
Section~\ref{sec:approach} presents the design of Barista;
Section~\ref{sec:casestudies} evaluates the Barista resource allocator for a prototypical case study;
and finally, Section~\ref{sec:conclusion} presents concluding remarks alluding to future directions.


\section{BACKGROUND AND RELATED WORK }
\label{sec:survey}

This section provides an overview of the literature along the dimensions of deep learning-based prediction services, infrastructure elasticity, serverless computing and workload forecasting, all of which are critical for the success of the presented work on Barista.

\subsection{ Deep Learning-based Prediction Services}
Pre-trained models based on deep learning techniques are increasingly being used in prediction analytics services.   In this approach, all the learned internal parameters are stored in the form of a vector of scores for each category along with their weights in a pre-trained deep learning model of desired accuracy~\cite{lecun2015deep}. Once the models are trained, these prediction models are seamlessly integrated into applications to predict outcomes based on
new input data.

\subsection{ Serverless Computing}
Serverless computing focuses on providing zero administration by automating deployment and management tasks.  In this paradigm, the responsibility of deployment and management is delegated to another entity, which could be the cloud infrastructure provider itself or a mediating entity. The execution platform leverages container technology to deploy and scale the prediction service components, which helps to minimize idle resource capacity~\cite{lloydserverless}. These features are beneficial to the design and deployment of parallelizable deep learning prediction services.

However, due to variation in workloads, the providers of prediction services are required to modify their resource requirements by monitoring the resources continuously~\cite{barve2018upsara}, and that reactive approach can often violate the SLOs. Barista intelligently and efficiently manages the containerized allocation based on resource estimation by workload forecasting and profiling the execution time of prediction services.

The MxNet deep learning framework~\cite{ishakian2017serving} shows the feasibility of using serverless computing AWS Lambda framework~\cite{lambda}. There are several efforts~\cite{ granchelli2017towards, bhattacharjee2018model, binz2013opentosca,di2015automatic, bhattacharjee2018wip} to deploy and orchestrate VMs or containers dynamically with all software dependencies. Barista's focus is orthogonal to these efforts; instead, it is to trigger the deployment process ahead of time so that the system can handle the workload surges by utilizing the afore-mentioned efforts.

\subsection{ Dynamic Infrastructure Elasticity}
Most state-of-the-art technology and research strategies to \emph{horizontally} or \emph{vertically} scale the resources are heuristics-driven or rule-based and have custom triggers.  Selecting optimal cloud configurations is an NP-hard problem, and various models are presented based on heuristics ~\cite{aldhalaan2015near,shekhar2017indices,brogi2017qos}.
In Barista, we consider leasing VMs from the cloud provider to meet the latency bounds by relying on time series forecasting of the incoming workloads. We propose an efficient heuristic to select configuration types to guarantee bounded prediction latency while minimizing the cost.

\begin{figure*}
	\centering
	\includegraphics[width=\textwidth]{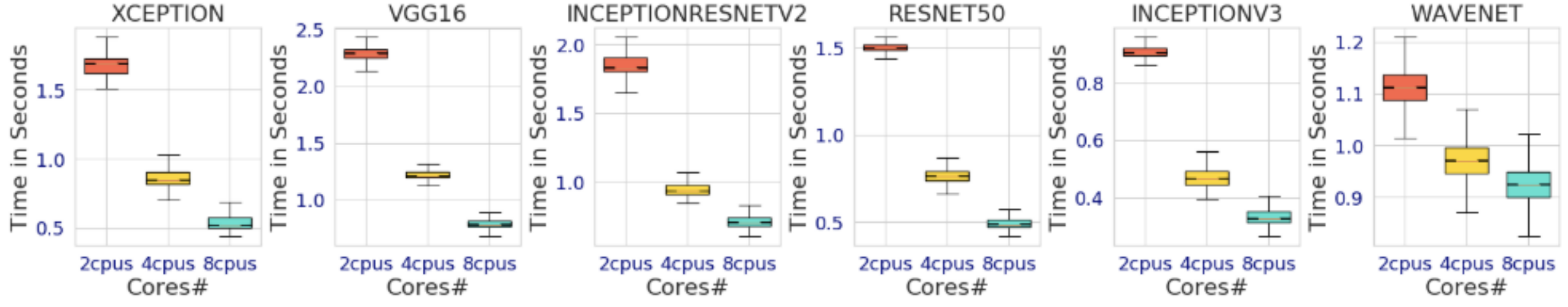}
	\caption{Box plots of prediction times for different deep learning pre-trained models on different numbers of CPU cores (2, 4 and 8).}
	\label{Fig:modeltime}
\end{figure*}

Swayam~\cite{gujarati2017swayam} presented a short-term predictive provisioning model to guarantee SLO while minimizing resource waste.  However, they only consider horizontal scaling by allocating more backend containers from the resource pool.  Model loading time for deep learning models is significant especially when the container is in cold state~\cite{ishakian2017serving}.  In contrast, Barista proactively considers infrastructure provisioning time to scale the system and also allows vertical resource scaling.

\emph{Vertical elasticity} adds flexibility as it eliminates the overhead in starting a new VM and loading the service model. Prior efforts to scale the CPU resources vertically appear in~\cite{lakew2014towards,kalyvianaki2014adaptive} including an approach that uses the discrete-time feedback controller leveraging MAPE-K loop for containerized applications~\cite{al2017autonomic}. Barista uses an efficient, proactive method to trigger the scaling of resources horizontally while relying on vertical scaling reactively to allocate and de-allocate CPU cores for model correction when our estimation model cannot predict accurately. Our reactive approach can also handle sudden workload spikes within a threshold.

\subsection{ Workload Forecasting}
Workload forecasting is indispensable for service providers to anticipate changes in resource demands and make proactive resource allocation decisions.  Various forecasting methods based on time series analysis are described in~\cite{herbst2014self}.
In AGILE~\cite{nguyen2013agile}, a resource prediction algorithm is proposed to scale up the server pool by renting the VMs from the cloud providers \emph{a priori} to guarantee the SLOs of the services. Similarly, Dejavu~\cite{vasic2012dejavu} and Bubble-Flux~\cite{yang2013bubble} proposed self-adaptive resource management algorithms, which leverage workload prediction and application performance characterization to predict resource requirements.  These efforts employ a linear model for workload prediction which often results in low-quality forecasts with high uncertainty.

Several non-linear methodologies  based on  Support Vector Machine~\cite{nikravesh2017autonomic}, Error Correction
Neural Network (ECNN)~\cite{islam2012empirical}, Gaussian processes~\cite{sheikh2011bayesian, shekhar2018performance} are proposed to predict workloads.  However, these models fail to capture longer-term trends which are characteristics of cloud-hosted services~\cite{taylor2017forecasting}. In~\cite{taylor2017forecasting}, a hybrid model called \emph{Prophet} is proposed for forecasting workloads by combining linear/logistic trend models with a Fourier series-based seasonality model. According to the authors, Prophet is easier to use than the widely used ARIMA models~\cite{al2017autonomic,calheiros2015workload}  as it handles missing values automatically. Moreover, the ARIMA models generally struggle to produce good quality forecasts as it lacks seasonality detection.

In general, workload forecasting methods tend to lack feedback to update predictions based on recent performance.  Therefore, Barista extends Prophet with a non-linear decision-based model that modifies the forecast according to previous prediction errors. Barista 
workload forecasting model estimates the resource requirement and proactively scales the infrastructure to guarantee application SLOs.


\section{SYSTEM MODEL AND PROBLEM DESCRIPTION}
\label{sec:problem}
This section first describes the infrastructure model and assumptions, and then presents the problem formulation, which has two subproblems. First, given the SLO of the service, the properties of the pre-trained model and the costs of VM configurations, we determine the cost-effective VM types by solving an optimization problem to meet the SLO (Section~\ref{sec:static}). We then consider the dynamic management of VM and container resources through workload forecasting and infrastructure elasticity (Section~\ref{sec:dynamic}).

\subsection{Infrastructure Model and Assumptions}\label{sec:model}
To explore the \emph{serverless capabilities}, we assume that the deep learning pre-trained model for a predictive analytics service is encapsulated inside containers which are executed in a cluster of VMs. All the requests to the service are assumed to be \emph{homogeneous}, i.e., they execute the same prediction model, and the service is stateless. Since deep learning models are generally compute-intensive, they benefit from executing on multiple cores.  We validate this claim in Figure~\ref{Fig:modeltime}, which shows the range of prediction time latencies for several pre-trained deep-learning models on a VM hosted on AMD Opteron 2300 (Gen 3 Class Opteron) physical machine with different numbers of assigned CPU cores, demonstrating good speedup behaviors. The results are obtained by running 10,000 trial executions in isolation for each model. 

The VM lifecycle in cloud infrastructure for service deployment and management is considered as follows.

\begin{enumerate}[label=\protect\circled{\arabic*}]
	\item \texttt{VM Cold}: VM has not been deployed.
	\item \texttt{VM Warm}:  VM is deployed, but the container inside the VM has not been downloaded.
	\item \texttt{Container Cold}: the container is downloaded, but the pre-trained deep-learning model has not been loaded into the container's memory.
	\item \texttt{Container Warm}: the deep-learning model is loaded, and the container is ready to serve the prediction requests.
\end{enumerate}

Finally, we assume that once a VM is deployed, it is leased from the cloud provider for a minimum duration of $\tau_{vm}$ time. During this time, the deployment cost is paid for even if the VM is not used (because we do not scale down the system immediately). This could happen when the prediction model is unloaded from the container's memory during lightly-loaded periods so that the VM could serve other batch jobs in the background (e.g., deep analytics applications).

\subsection{VM Flavor Selection and Initial Deployment}\label{sec:static}

We first consider the static problem of serving a fixed set of requests that execute a deep-learning prediction model by finding the most cost-effective VM flavor type. Let $\lambda$ denote the constraint specified by the SLO of the model regarding its execution latency\footnote{Depending on the SLO, the execution latency can be flexibly defined based on, e.g., worst-case latency, $x$-percentile latency. In this paper, we consider the 95th percentile latency.}, and let $t_p$ denote the latency when the model is executed on $p$ CPU cores. Further, let $min\_mem$ denote the minimum amount of memory required to run a prediction model. 

To serve the prediction requests, we need to deploy a set of VMs from the cloud provider that offers a collection of $m$ possible configurations (flavors), denoted as $\{vm_1, vm_2, \dots, vm_m\}$. We consider three parameters to specify each configuration $vm_i$: \circled{1} $p_i$, the number of available cores; \circled{2} $mem_i$, the memory capacity; and \circled{3} $cost_i$, the cost of deployment. In particular, the cost includes both the running cost and the management cost of deploying the VM. For each configuration $vm_i$, suppose $\alpha_i \in \{0, 1, 2, \dots, \}$ number of VMs are deployed. Then, the total deployment cost is given by $total\_cost = \sum_{i=1}^{m} \alpha_i \cdot cost_i$. While each deployed VM is assumed to serve only one request at a time (because each request benefits from consuming all the cores as the prediction service is highly-parallelizable), it could serve multiple requests one after another. In this case, the request that is served later in the pipeline needs to wait for the preceding requests to be first completed, which will delay its prediction time.

The optimization problem concerns deploying a set of VMs, i.e., to choose $\alpha_i$'s for all $i = 1,\dots, m$, so that the total cost is minimized while the SLOs of all the requests can be satisfied (i.e., with latency not larger than $\lambda$). Section \ref{sec:vm_deploy} presents our VM deployment solution.

\subsection{Dynamic Resource Provisioning via Workload Forecasting and
  Infrastructure Elasticity}\label{sec:dynamic}

The resources must be provisioned dynamically to meet the SLOs under workload variations. Since the reactive approach can be detrimental to response times, we use a proactive approach to handle the variation of the workload by forecasting the future service demands based on historical workload patterns of a deep-learning prediction model.

\begin{figure}[htb]
	\centering
	\includegraphics[width= 0.7\columnwidth]{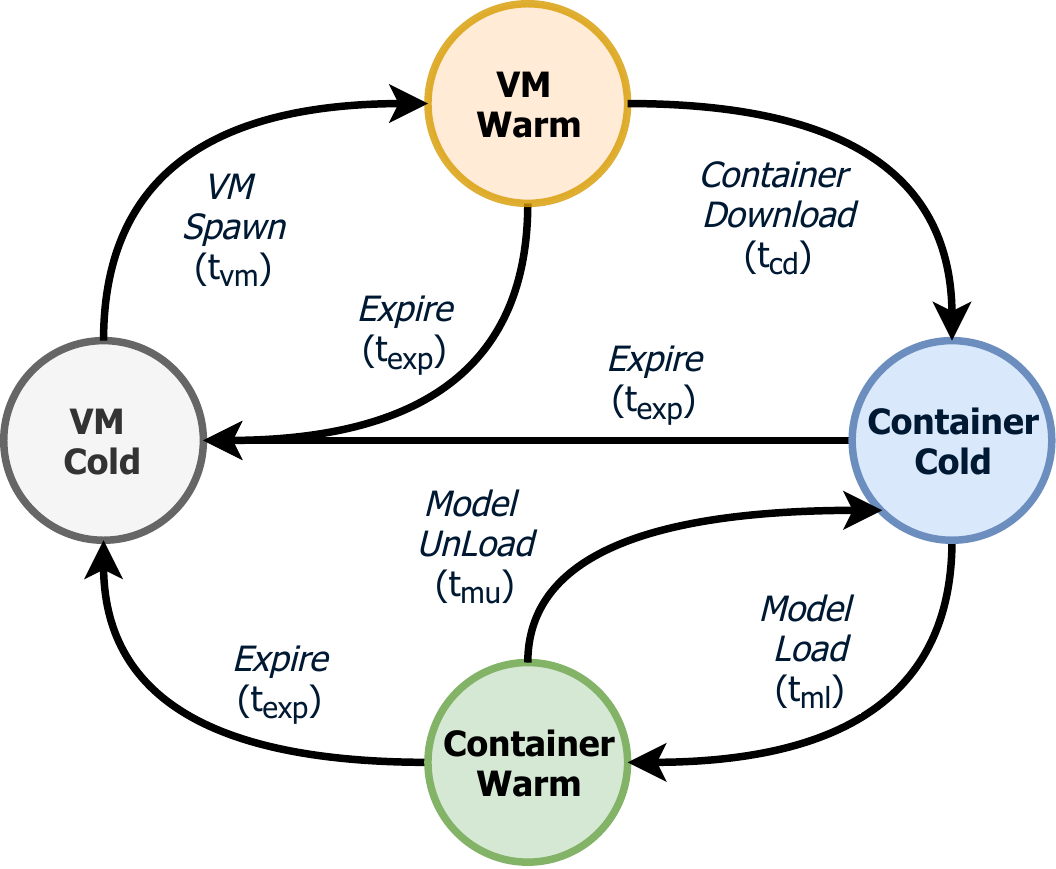}
	\caption{An abstract state machine showing different states and transitions associated with a life cycle of a VM in cloud infrastructure. Edges are labeled with actions and time duration to complete the state transition.}
	\label{Fig:vmstate}
\end{figure}

With the forecasted future workload (i.e., number of service requests), the VM deployment decision (as described in Section \ref{sec:static}) must also be adapted accordingly. Horizontal scaling~\cite{lorido2014review} is a promising approach to provision the resources dynamically. To that end, we exploit the four cloud infrastructure states described earlier. Figure \ref{Fig:vmstate} illustrates the actions needed to transition between the states. Each action incurs a state transition time.  Specifically, we denote the VM deployment time by $t_{vm}$, the container service download time by $t_{cd}$, and the pre-trained model loading time by $t_{ml}$.\footnote{The time required to unload the model from memory, denoted by $t_{mu}$, is negligible and thus not considered. The time taken to move from any state to \texttt{VM Cold} is denoted by $t_{exp}$. This duration does not impact the resource manager logic and hence is also ignored.}

\begin{figure}[tbh]
  \centering
  \includegraphics[width= \columnwidth]{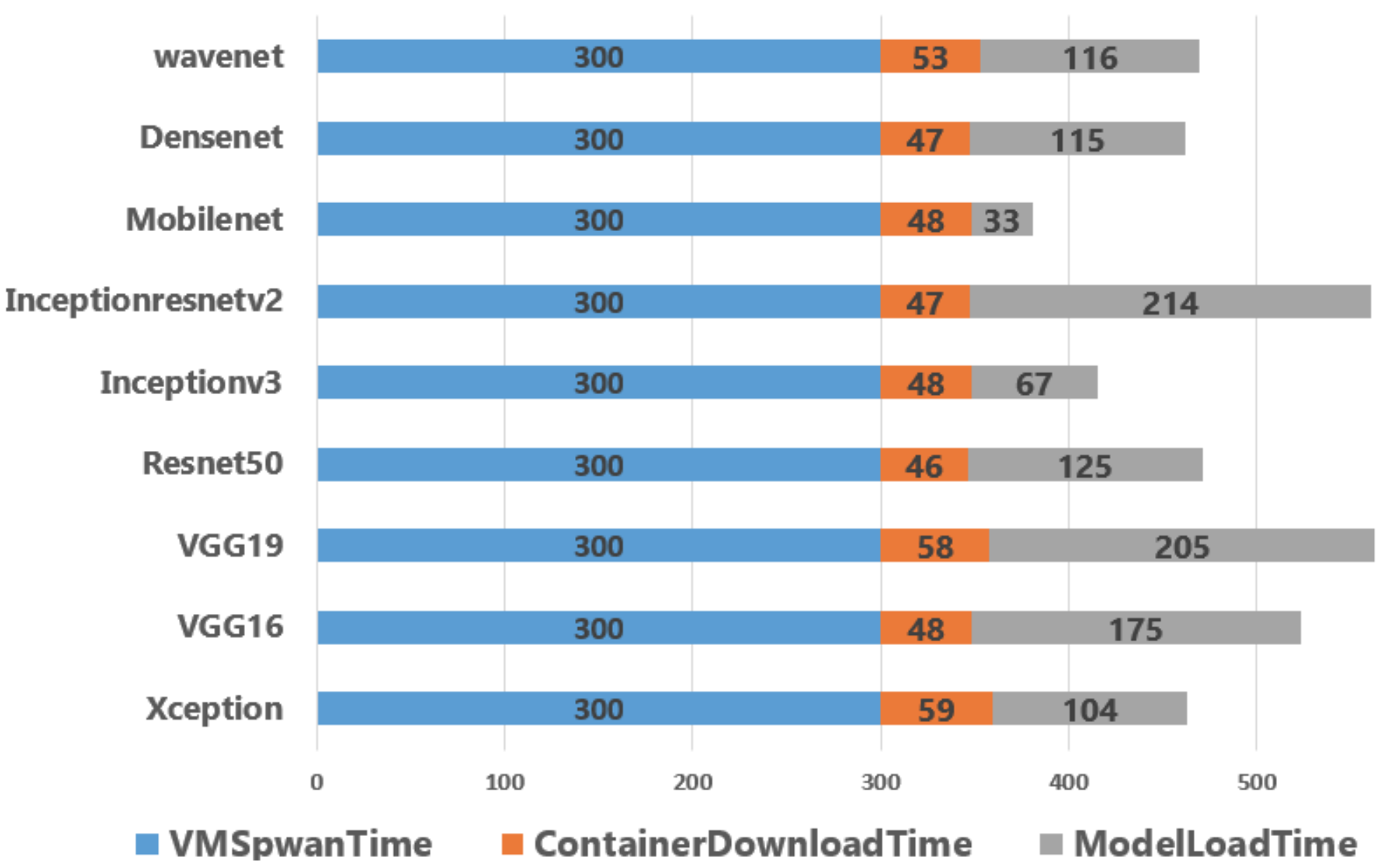}
  \caption{The setup times (in seconds) for different deep learning prediction models as per our experiment. The blue bars show the VM deployment time ($t_{vm}$), the oranges bar show the specific pre-trained model container download time ($t_{cd}$), and the grey bars show the prediction model loading time ($t_{ml}$).}
  \label{Fig:modelspwanTime}
\end{figure}

Figure~\ref{Fig:modelspwanTime} shows concrete timings for the different prediction services we tested on our experimental infrastructure. Further, we refer to the total time to set up the service as $t_{setup} = t_{vm} + t_{cd} + t_{ml}$. This motivates the need to forecast the future workload $t_{setup}^\prime = t_{setup} + t_{forecast}$ time ahead to guarantee the SLO with certain accuracy, where $t_{forecast}$ is the time taken to obtain the
forecast. The forecasting needs to be performed for $t_{setup}^\prime$ time into the future to account for the infrastructure setup time.

When the VM is not serving any requests, the infrastructure state transitions from \texttt{Container Warm} to \texttt{Container Cold}. Later on, when the load increases and the VM needs to serve requests again, the model will be reloaded. As a result, we also need to check the infrastructure state ahead of time to make decisions for downloading the container and loading the model if it is not already in the \texttt{Container Warm} state. We account for all of these times in meeting the SLOs, while avoiding excess over-provisioning of the infrastructure resources. Section \ref{provisioner} presents our combined solution to the dynamic resource management problem that incorporates infrastructure elasticity, workload forecasting, and VM deployment.


Moreover, if the workload forecaster over-estimates the workload, we allow the excess resources to be utilized by the low-priority batch jobs via vertical scaling. Co-locating various jobs on a server can cause performance interferences.  In our approach, we assume 20\% performance degradation (for the worst-case scenario based on our experiment) if a latency-sensitive prediction service is co-hosted with batch jobs.


\section{DESIGN \& IMPLEMENTATION of BARISTA}
\label{sec:approach}

In this section, we give the architectural insights of Barista by describing its various components. We also explain our solutions to the problems of static VM deployment and dynamic resource provisioning as mentioned in Section~\ref{sec:problem}.

\subsection{Architecture of Barista}

\begin{figure}[htb]
	\centering
	\includegraphics[width= 0.95\columnwidth]{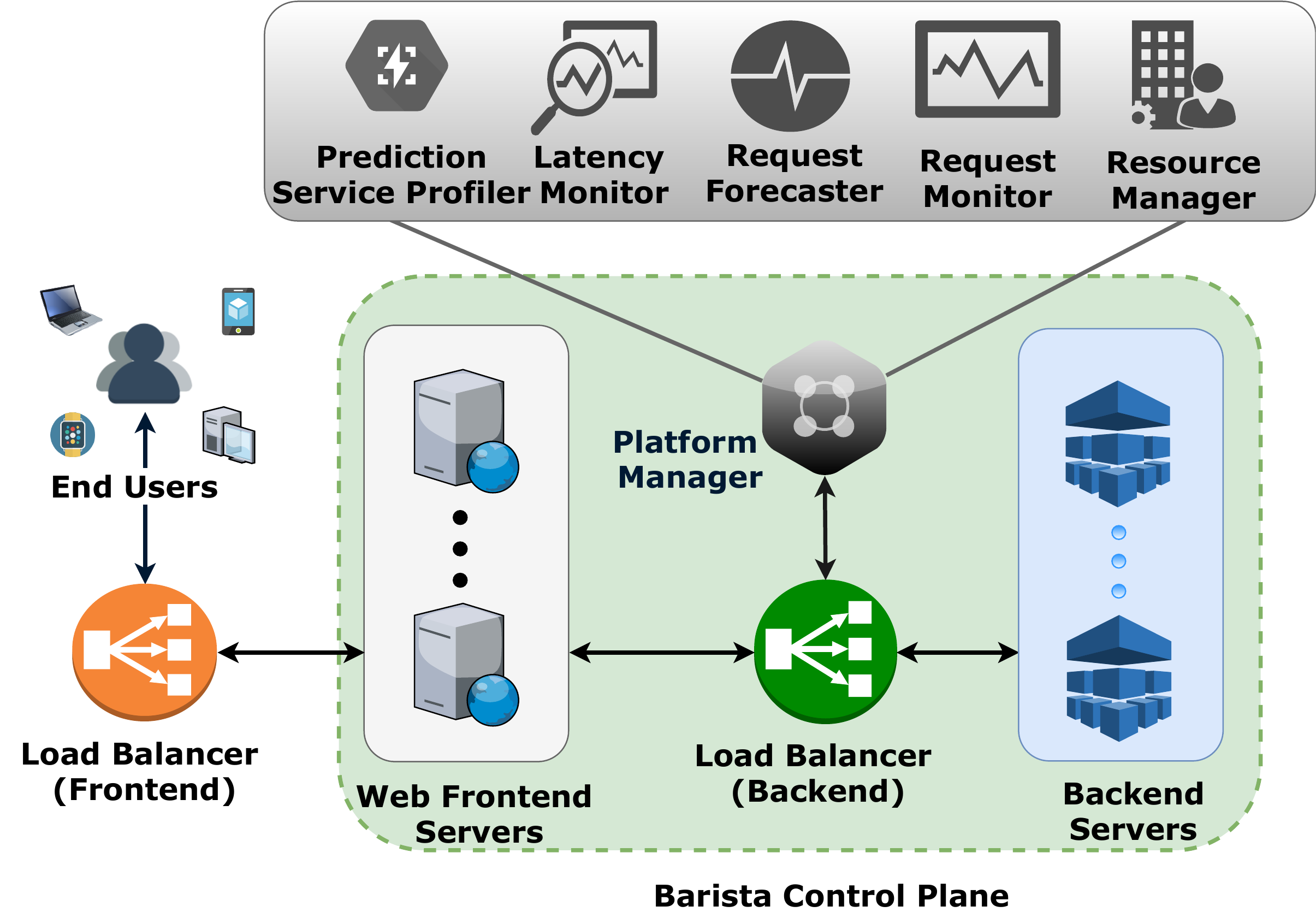}
	\caption{Architecture of Barista serving system.}
	\label{Fig:baristaArch}
\end{figure}

\begin{figure}[tbh]
	\centering
	\includegraphics[width=0.95\linewidth]{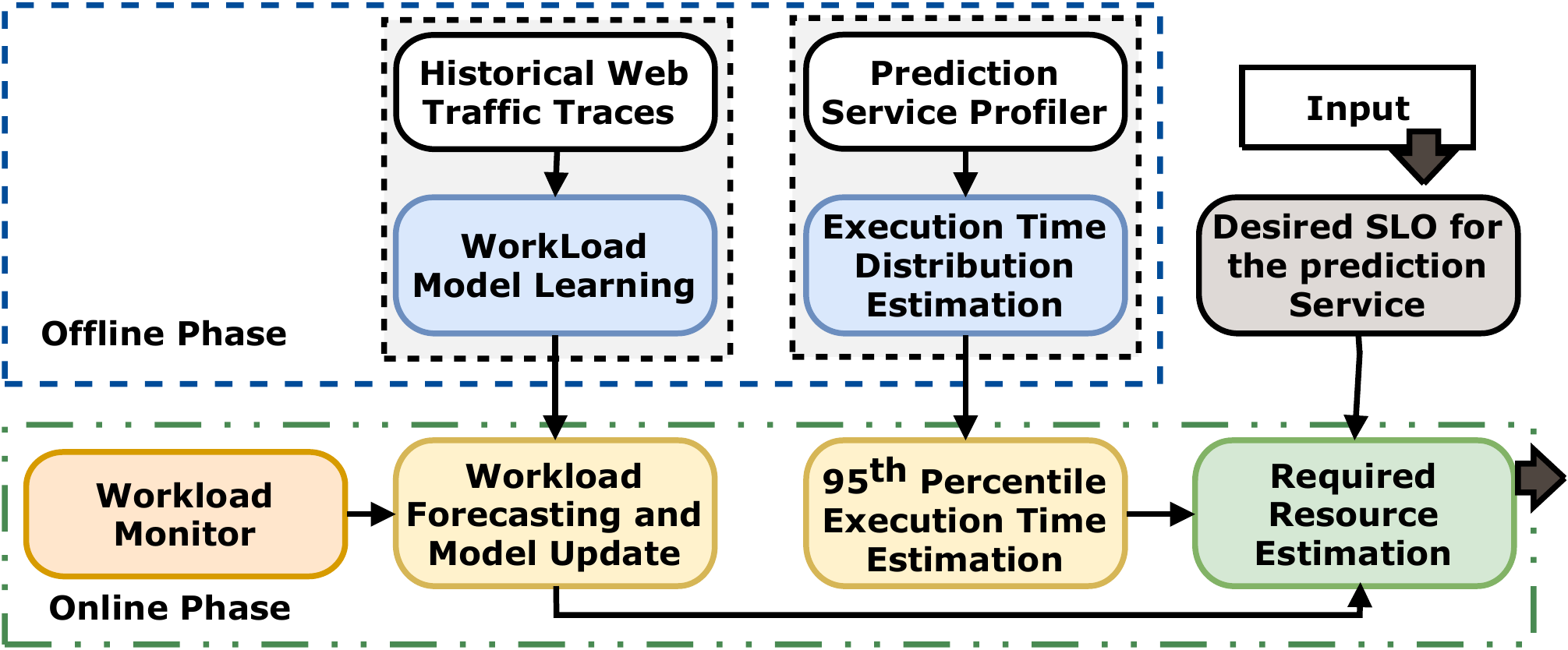}
	\caption{Data flow model of Barista platform manager.}
	\label{Fig:baristamodel}
\end{figure}

Barista architecture consists of a pool of frontend and backend servers, load balancers to distribute the requests, a platform manager to allocate and scale backends for different prediction services as shown in Figure~\ref{Fig:baristaArch}. Frontend servers are the virtual machines, which host the user interface, whereas backend servers host the containerized pre-trained deep learning model.   End users send their requests to the frontend load balancer, which redirects the requests to the frontend servers based on the \emph{round robin} policy.  Each frontend server forwards the request to the backend load balancer, which then redirects
the request to one of the backend servers assigned to serve the prediction query based on the \emph{least loaded connection} policy.  Each backend processes a single request at a time and gives the prediction result back. The platform manager is an integral part of Barista, which is responsible for dynamic provisioning of resources in cloud infrastructure by forecasting workload patterns and estimating execution time of various prediction
queries. The platform manager (as zoomed in Figure~\ref{Fig:baristaArch}) consists of a prediction service profiler, a request monitor, a request forecaster, a prediction latency monitor, and a resource manager. They are described as follows:

\begin{enumerate}[label=\protect\circled{\arabic*}]
    \item \textbf{Prediction Service Profiler:}  It profiles the execution time of a prediction service on different numbers of CPU cores and finds the best distribution (as shown in Figure~\ref{Fig:baristamodel}). This provides the 95th percentile latency estimate of the execution time for the prediction service based on the assigned number of cores.
	\item \textbf{Request Monitor:} It monitors and logs the number of aggregated incoming requests received every minute by the backend load balancer.
	\item \textbf{Request Forecaster:} It predicts the number of requests $t^\prime_{setup}$ time steps into the future based on the historical data. The forecaster also updates its model every minute based on the previous prediction errors and the actual data from the request monitor to diminish uncertainty.
	\item \textbf{Prediction Latency Monitor:} It monitors and logs the SLO violations for the incoming requests every five seconds. SLO is defined over the response time of the backend servers to a prediction query requested by the frontend servers.
	\item \textbf{Resource Manager:} It allocates the required number of virtual machines for the forecasted workloads and performs intelligent scaling based on resource estimation and provisioning strategies as discussed in Section~\ref{sec:vm_deploy}.
\end{enumerate}

The operation of the platform manager can be categorized into two phases, \textit{online} and \textit{offline}, as shown in Figure~\ref{Fig:baristamodel}. In the offline or design phase, the execution time of the deep learning model is profiled on different VM configurations followed by distribution estimation. Workload forecasting model is also trained in this phase. In the online or runtime phase, based on the output of workload forecaster and execution time estimator, resources are estimated and provisioned. The actual workload is also being monitored and stored, which is used to update the forecasting model based on the last five error predictions and rolling training window. The following subsections provide more details on these operations.



\vspace{-0.1in}


\subsection{ Execution Time Distribution Estimation }\label{thexec}
Inaccurate estimate of the execution time of a pre-trained deep learning model can result in erroneous output produced by the resource manager, which may lead to over- or under-provisioning of resources. Thus, in Barista, extensive offline profiling of different deep learning models is performed on various VM configurations. Figure~\ref{Fig:modeltime} shows our experiments on configurations involving 2, 4 and 8 CPU cores with required memory size, where each experiment contains 10,000 trails. The execution times are random in nature and follows an unknown distribution. In Barista, the resource manager uses the 95th percentile statistic of the execution time to provision resources. In order to accurately calculate the percentile values,  we remove sample bias and estimate the unknown distribution.

Barista prediction service profiler estimates the distribution using parametric methods based on Maximum Likelihood Estimation (MLE) for fitting different distributions and finding their respective unknown parameters. For quantifying the goodness of fit, we use one-sample Kolmogorov-Smirnov (K-S) test~\cite{wilcox2005kolmogorov} to rank different hypothesized distributions.  Given the cumulative distribution function $F _{0}$ of the hypothesized distribution and the empirical distribution function $F_{data}(x)$ of the observed dataset, the test statistic ($D_n$ ) can be calculated by Equation~(\ref{eq:KS}), where $sup_{x}$ is the \textit{supremum} of the set of distances and $n$ is the size of the data. According to~\cite{tucker1959}, if the sample comes from distribution $F_{0}(x)$, then $D_n$ converges to 0 almost surely in the limit when $n$ goes to infinity.
\begin{equation}
\label{eq:KS}
D_n = sup_{x} \vert F_{0}(x) - F_{data}(x) \vert
\end{equation}

\subsection{Workload Forecasting}\label{sec:forecast}

Barista uses an online rolling window based forecasting methodology to predict workload in order to allocate resources proactively. Request forecaster is composed of two main components: \circled{1} \emph{Forecaster}, which is responsible for modeling both periodic and non-periodic elements associated with time-varying workloads, and  \circled{2} \emph{Compensator}, which modifies the forecast produced by the first component according to the last five forecast errors. These two components are briefly described as follows:

\subsubsection{Forecaster}\label{sec:forecaster}
A time-varying workload can be composed of three main elements - \emph{trend}, \emph{seasonality} and \emph{holidays}~\cite{hastie1987generalized, taylor2017forecasting}, and they are combined as shown in Equation~(\ref{eq:PR}) below:
\begin{equation}
\label{eq:PR}
y(t) = g(t) + s(t) + h(t) + \epsilon_{t}
\end{equation}
where $g(t)$, $s(t)$, $h(t)$ model non-periodic changes, periodic changes (e.g., daily, weekly and yearly seasonality), and effects of holidays which occur on potentially irregular schedules over one or more days, and $\epsilon_{t}$ represents noise.

The trend function ($g(t)$) models how the workload has grown in the past and how it is expected to continue growing. Modeling web traffic is often similar to population growth in natural ecosystem, where there is a non-linear growth that saturates at carrying capacity~\cite{taylor2017forecasting}. This kind of growth is typically modeled using the logistic function shown in Equation~(\ref{eq:TR}), where $C$ is the carrying capacity, $k$ is the growth rate, and $m$ is an offset parameter.
\begin{equation}
\label{eq:TR}
g(t) = \frac{C}{1 + \exp(-k(t-m))}
\end{equation}

The seasonality function ($s(t)$) models multi-period seasonality that repeats after a certain period. For instance, a five day work week can produce effects on a time series that repeat each week. A standard Fourier series, as shown in Equation~(\ref{eq:FS}), is used to provide a flexible model of periodic effects~\cite{harvey1993structural}, where $P$ is the
expected time series period and $N$ is the order.
\begin{equation}
\label{eq:FS}
s(t) = \frac{1}{2} \, a_{0} + \sum_{n=1}^{N} \left[
a_{n}\,\boldsymbol{\cos} (\frac{2 \pi n t}{P}) + b_{n} \,\boldsymbol{\sin} (\frac{2 \pi n t}{P}) \right]
\end{equation}

The holiday function ($h(t)$) models the predictable variations in workload caused due to holidays. However, these variations do not follow a periodic pattern, so their effects are not well modeled by a cycle. The holidays are
added in the form of a list and are assumed independent of each other. An indicator function is added for each holiday that shows the effect of a given holiday on the forecast. Barista leverages Prophet~\cite{taylor2017forecasting} for implementing Forecaster.

\subsubsection{Compensator}\label{sec:compensator}
This component adjusts the output of the forecaster based on the past forecast errors. It can be modeled as a transformation function $c$, which changes the output of the forecaster $y$ based on the errors from the last $m$ forecasts $E=\{e_1, e_2, \dots, e_m\}$, as shown in Equation~(\ref{eq:COMP}) below:
\begin{equation}
\label{eq:COMP}
y'=c(y, y_{upp}, y_{low}, E)
\end{equation}
where $y_{upp}$ and $y_{low}$ are the upper and lower estimation bounds of $y$. The transformation can be learned using data-driven methods. In Barista, we use H2O's AutoML framework~\cite{automl} to find the best hyper-parameter tuned algorithm.

\subsection{Resource Estimation}\label{sec:vm_deploy}
\emph{Resource estimation} is one of the two main tasks performed by the resource manager. Depending upon the forecasted value, the SLO and the service type, the resource manager solves the static VM deployment problem described in Section \ref{sec:static}.  Due to the NP-hardness of the problem, this subsection presents a greedy heuristic to perform static VM deployment.

For each VM configuration $vm_i$, we can compute the number of requests $n\_req_i$ it is able to serve for a deep learning prediction service while meeting the SLOs:
\begin{align*}
n\_req_i = \begin{cases}\lfloor \frac{\lambda}{t_{p_i}} \rfloor, & \text{if } mem_i \ge min\_mem  \\ 0, & \text{otherwise} \end{cases}
\end{align*}
Recall that $\lambda$ is the model's SLO timing constraint, $min\_mem$ is the model's minimum memory requirement, $t_{p_i}$ is the latency to serve each request of the model using configuration $vm_i$ with $p_i$ CPU cores, and $mem_i$ is amount of memory available in $vm_i$.
We can then define the \emph{cost per request} for each configuration $vm_i$ as follows:
\begin{align*}
cpr_i = \frac{cost_i}{n\_req_i}
\end{align*}

Let $i^*$ denote the index of the VM configuration with the minimum cost per request, i.e., $cpr_{i^*} = \min_{i = 1\dots m} \{cpr_i\}$. Clearly, given an estimated workload $y'$ from the output of Equation~(\ref{eq:COMP}), an optimal rational solution will deploy $\alpha^* = \frac{y'}{n\_req_{i^*}}$ VMs of configuration $vm_{i^*}$ and incurs a total cost:
\begin{align}\label{eq:optimalcost}
total\_cost^* = \frac{y'}{n\_req_{i^*}}\cdot cost_{i^*}
\end{align}
To find the optimal integral solution is unfortunately NP-hard (via a simple reduction from the knapsack or the subset sum problem). 
However, Equation (\ref{eq:optimalcost}) nevertheless serves as a lower bound on the optimal total cost.

To solve the integral problem, our greedy algorithm also chooses a single configuration $vm_{i^*}$ that has the minimum cost per request while breaking ties by selecting the configuration with a smaller deployment cost. Thus, it deploys $\alpha = \lceil \frac{y'}{n\_req_{i^*}} \rceil$ VMs of configuration $vm_{i^*}$ for serving $y'$ requests, and incurs a total cost that satisfies:
\begin{align}\label{eq:greedycost}
total\_cost &= \lceil \frac{y'}{n\_req_{i^*}} \rceil \cdot cost_{i^*} \nonumber  \\
&< \big(\frac{y'}{n\_req_{i^*}} + 1 \big) \cdot cost_{i^*} \nonumber  \\
&= total\_cost^* + cost_{i^*}  
\end{align}
Equation~(\ref{eq:greedycost}) shows that the total cost of the greedy algorithm is no more than the optimal cost plus an additive factor $cost_{i^*}$. When serving a large number of requests, the incurred cost is expected to be close to the optimal. Furthermore, the algorithm always deploys VMs from the same configuration regardless of the number of requests to be served. This makes it an attractive solution for handling dynamic workload variations without switching between different VM configurations. The complete algorithm is illustrated in Algorithm~\ref{resourcelogic}.

%
%
%
%
\begin{algorithm}
	\small
	\caption{Resource Estimation}\label{resourcelogic}
	\SetAlgoLined
	Initialize: $i^* \leftarrow 0$, $cpr^* \leftarrow \infty$, $cost^* \leftarrow \infty$, $n\_req^* \leftarrow 0$ \\
	\For{$i=1$ to $m$ }{
		$t_{p_i}$ $ \leftarrow $ getExecutionTime$(vm_i,model)$  \\
        $mem_i \leftarrow \text{getMemory}(vm_i)$ \\
        $cost_i \leftarrow \text{getCost}(vm_i)$ \\
		\If{$mem_i$ $\ge$ $min\_mem$}{
			$n\_req_i \leftarrow \lfloor \frac{\lambda}{t_{p_i}} \rfloor $  \\
			$cpr_i$ = $\frac{cost_i}{n\_req_i}$ \Comment Cost per request \\
			\uIf{$cpr_i < cpr^*$}{
                $i^* \leftarrow i$\\
                $cpr^* \leftarrow cpr_i$ \\
				$n\_req^* \leftarrow n\_req_i$ \\
                $cost^* \leftarrow cost_i$}
            \ElseIf{$cpr_i = cpr^*$ \& $cost_i < cost^*$}{
                $i^* \leftarrow i$\\
                $n\_req^* \leftarrow n\_req_i$ \\
                $cost^* \leftarrow cost_i$
            }
		}
	}
	Deploy $\alpha \leftarrow \lceil \frac{y'}{n\_req^*} \rceil$ VMs of configuration $vm_{i^*}$\\
\end{algorithm}
\vspace{-0.1in}

\subsection{Resource Provisioner}\label{provisioner}

 \emph{Resource provisioning} is another main process running inside the Barista resource manager. It is implemented as a daemon process that is invoked at a fixed interval of time as shown in Algorithm~\ref{allocationlogic}. On each invocation, the resource manager obtains a workload forecast $t_{setup}^\prime$ timesteps into the future [Line 4]. Depending upon the forecasted workload, using Algorithm~\ref{resourcelogic}, a required number of VMs (i.e., $\alpha$) from a best VM configuration can be calculated.

\begin{algorithm}
	\small
	\caption{Resource Provisioning}\label{allocationlogic}
	\SetAlgoLined
	Initialize: $Flag \leftarrow True$, $\alpha \leftarrow 0$, $n\_req_{i^*} \leftarrow 0$, $ i^* \leftarrow 0 $, $params \leftarrow \{model, \lambda, min\_mem, mem_i, p_i, cost_i, \forall i =1 \dots m\}$, $prevStepVMCount \leftarrow 0$, $scaledVMs \leftarrow \emptyset$ \\
	\While{$True$}{
		$t \leftarrow $ GetCurrentTime()\\
		$y' \leftarrow $ GetForecast$(t, t'_{setup})$\\
		\uIf{$Flag$}{
			$\alpha, i^*, n\_req_{i^*} \leftarrow $ ResourceEstimation$(y', params)$\\
			$Flag \leftarrow False$\\
		}
		\Else{
            \vspace{-0.05in}
			$\alpha \leftarrow \lceil \frac{y'}{n\_req_{i^*}} \rceil$
		}
		$expireVMCount \leftarrow $ GetExpireVMCount$(t + t'_{setup})$\\
		$\delta \leftarrow ( \alpha - prevStepVMCount ) - expireVMCount$ \\
		
		\uIf { $\delta > 0$}{
			\For{$i=1$ to $\delta$}{
				$IP \leftarrow $DeployVM$(i^*)$\\
				AddContDwnldRegistry$(t+t_{vm}, IP, model)$\\
				AddMdlLoadRegistry$(t+t_{vm}+t_{cd}, IP, model)$ \\
				AddVMExpireRegistry$(t+\tau_{vm}, IP, model)$ \\
			}
			HorizontalScaleUp$(sizeof(scaledVMs), scaledVMs)$\\
		}\Else{
			$\delta^\prime \leftarrow \delta + sizeof(scaledVMs)$\\
			\uIf {$\delta^\prime \geq 0$}{
				HorizontalScaleUp$(\delta^\prime, scaledVMs)$\\
			}
			\Else{
				HorizontalScaleDown$(|\delta^\prime|, scaledVMs)$\\
			}
		}
		$C \leftarrow  $ CheckContDwnldRegistry($t$)\\
		$M \leftarrow  $ CheckMdlLoadRegistry($t$) \\
		$E \leftarrow  $ CheckVMExpireRegistry($t$) \\
		
		\ForEach{$c \in C$}{
			DownloadContainer($c.IP, c.model$)\\
		}
		\ForEach{$m \in M$}{
			LoadModel($m.IP, m.model$) \\
		}
		\ForEach{$e \in E$}{
			ModelUnload($e.IP, e.model$)\\
			TerminateVM($e.IP$)}
		
		$prevStepVMCount \leftarrow \alpha $\\
		LoadBalancerUpdate() \\
		WakeUpAtNextTick()\\
	}	
\end{algorithm}

Since the heuristic presented in Section~\ref{sec:vm_deploy} depends upon the SLO and cost per request, the best VM configuration will remain fixed as long as these two factors are unaltered. Hence, the best VM configuration index and the maximum number of requests served per VM are calculated once and stored in variables $i^*$ and $n\_req^*$, respectively [Lines 5-10].  After obtaining the required number of VMs, $\alpha$, difference between $\alpha$ and $prevStepVMCount$ is calculated to find the net VMs needed at timestep $t+t_{setup}^\prime$. From this difference, $expireVMCount$ (the number of VMs that will expire at time $t+t_{setup}^\prime$) is subtracted to compensate for the VMs that will become unavailable due to lease expiration. The final difference value is referred to as $\delta$ [Lines 11-12].

A positive value of $\delta$ implies more VMs are required than the current availability at time $t+t_{setup}^\prime$. In this case, $\delta$ number of new VMs are deployed, and container download, model upload, and VM expiration registers are updated [Lines 14-19]. Remember, if the VM is not serving the prediction task (see Figure~\ref{Fig:vmstate}), we first unload the prediction model, and allocate other batch jobs in the VM. These VMs are in \texttt{Container Cold} state before expiring. Now, if requests surge, all the previously allocated VMs, which are in \texttt{Container Cold} state, are re-instantiated for serving the prediction service [Line 20]. \emph{HorizontalScaleUp} checks the current infrastructure state of the VMs, which are in \texttt{Container Cold} state, and load the prediction model.
However, in the other case where $\delta$ is negative, the VMs can be horizontally scaled up or down depending upon the expression $\delta' \leftarrow \delta + sizeof(scaledVMs)$ [Line 22]. If $\delta^\prime$ is positive, scaling up is performed and vice-a-versa [Lines 23-27]. \emph{HorizontalScaleDown} unloads the prediction model and stops the prediction serving container, however, in \texttt{Container Cold} state the VM is part of prediction task cluster so that it can be recalled if necessary. The Resource manager also performs container download, model load and VM termination actions for previous forecasts at a current time step (i.e., $t$), followed by terminating VMs whose lease is expiring at $t$ [Lines 29-41]. In the end, the resource manager saves the forecast, updates load balancer for newly deployed VMs, and sleeps till the next tick [Lines 42-44].

\begin{figure*}
	\centering
	\includegraphics[width=\textwidth]{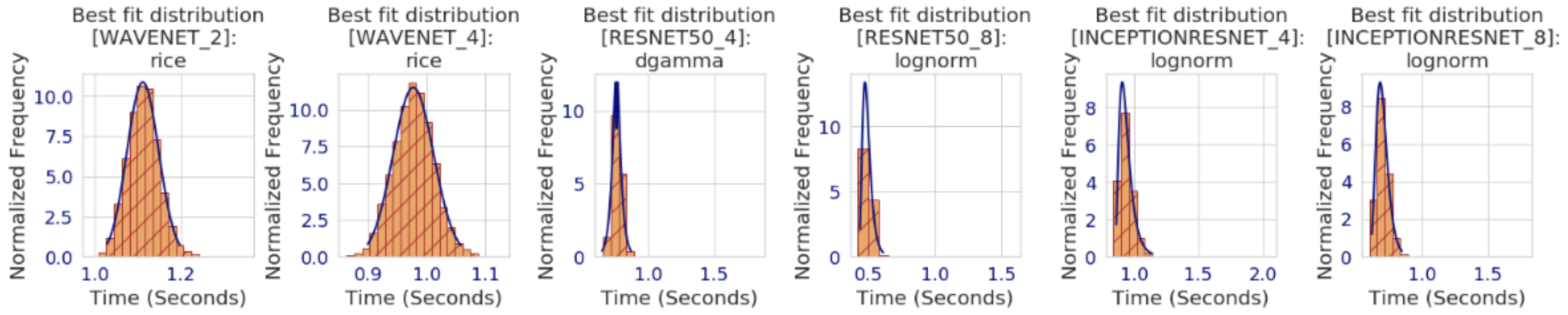}\vskip -2mm
	\caption{Top ranked distribution that describes the variation in the sample data. The distribution (\color{blue}blue\color{black}) is plotted on top of the histograms (\color{orange}orange\color{black}) of observations. }
	\label{Fig:histmodel}\vskip -2mm
	\vspace{-0.1in}
\end{figure*}

We vertically scale down the number of cores of a particular container if we meet the SLO with some threshold margin, and share the cores with the batch jobs.  Vertical scaling is also helpful to allocate more resources for sudden workload surges. Moreover, if the resource estimator over-estimates the resources and whenever our prediction services can de-allocate cores while maintaining the SLO, Barista frees up cores, and if we miss any SLO, Barista will increase the number of cores immediately if more cores are available. We de-allocate one core at a time to minimize latency miss, and we double the number of cores (within maximum core limits of the VM) for the prediction service if there is any SLO miss.


\section{EVALUATION}
\label{sec:casestudies}
In this section, we present experimental validation for the different
phases of our Barista framework.

\subsection {Experiment Setup}
Our testbed comprises an OpenStack Cloud Management system running on a cluster of AMD Opteron 2300 (Gen 3 Class Opteron) physical machines. We emulated the VM configurations as per the Amazon EC2 pricing model\footnote{We considered 47 different VM configurations with 2, 4 and 8 cores. Any GPU-based or SSD-based configuration was not considered.}~\cite{amazonec2}. We used the AWS pricing model to emulate our cost model\footnote{We did not run the experiments on Amazon cloud because of monetary constraints.}. We employed DockerSwarm~\cite{swarm} as our container management service on top of the VMs. HAProxy (\url{http://www.haproxy.org}) was used as our frontend and backend load balancers. We built a NodeJS-based frontend web application to relay the query to the backend predictive analytics service.

\subsection{Predicting Execution Time of Predictive Analytics Services} \label{evalexec}

In this experiment, we considered two different kinds of predictive analytics applications: image recognition and speech recognition, based on six different pre-trained deep learning models: Xception, VGG16, InceptionV3, Resnet50, InceptionResnetV2, and Wavenet (see Figure~\ref{Fig:modeltime}). All these models were profiled on OpenStack VMs of variable numbers of CPU cores.
Based on the data generated from profiling these models, the best distribution was estimated from a list of available probability distributions. All the empirical distributions closely resemble the hypothesized distributions. For example, the best-fit distributions for the Wavenet service on two and four cores, for the Resnet50 service on four and eight cores, and for the InceptionResnetV2 service on four and eight cores are shown in Figure~\ref{Fig:histmodel}. Based on the best-fit distributions, we calculated the 95th percentile latency for each service.

\subsection{Workload Forecasting }
Two different time series datasets were used to emulate a realistic workload for predictive analytics services. The first dataset was collected and published by NYC Taxi and Limousine Commission~\cite{nycdata}. We processed the data to extract the number of cab requests generated every minute based on the pick-up and drop-off dates and times~\cite{nycdata}. This dataset provides an appropriate workload for a speech recognition component in the ride-sharing application to request a ride. The second dataset contains data on the number and types of vehicles that entered from each entry point on the toll section of the Thruway with their exit points~\cite{nycthru}. This dataset can be used to represent a real-world workload of an image recognition based predictive analytics service that aims to automatically detect the license plate number of the entering or exiting car from a toll plaza. We processed the data to extract the total number of cars entering a toll plaza every minute.

A total of 10,000 data-points from each dataset was used in our study. We utilized 6000, 500 and 2500 data points for training, validation and testing the Prophet-based Forecaster model in both datasets, respectively. We performed \emph{hyper-parameter tuning} for the Fourier series order \textit{N} by iterating over five different values $\{10, 15, 20, 25, 30\}$, and with different sizes of the training window $W\in \{4000, 5000, 6000\}$. Out of the 15 possibilities, the configurations with $N=30, W=6000$ and $N=20, W=6000$ produced the least absolute percentage error (95th percentile) for the first and second datasets, respectively. The mean absolute error and absolute percentage error (95th percentile) for the first and second datasets are (27.66, 29$\%$) and (27.84, 30.26$\%$), respectively.

\begin{figure}[htb]
	\begin{minipage}{\columnwidth}
		\centering
		\includegraphics[width=0.8\linewidth]{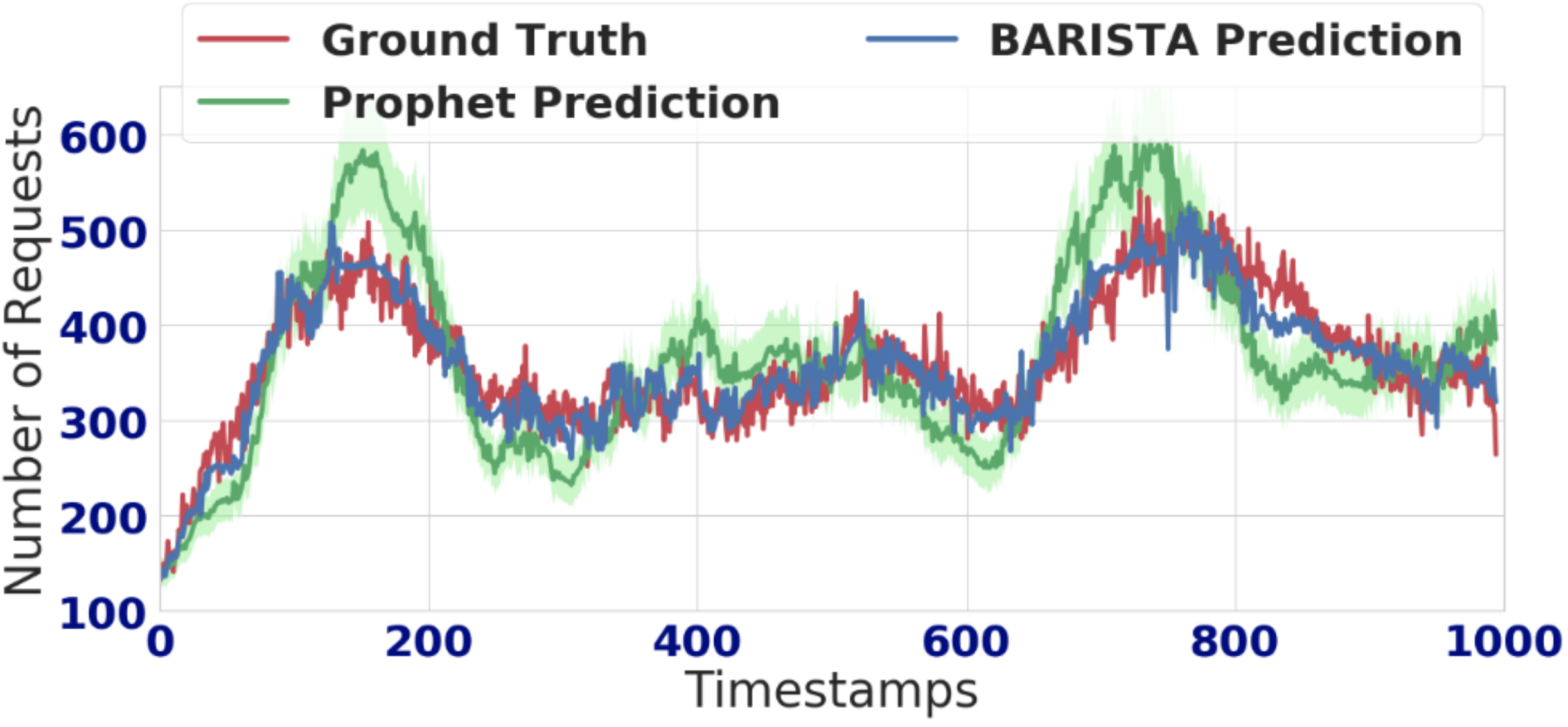}		\vskip -2mm
		\captionsetup{width=0.95\linewidth}
		\caption{Performance comparison of Barista (\color{black!20!blue}blue\color{black}) and Prophet (\color{black!50!green}green\color{black}) along with ground truth (first dataset) (\color{black!30!red}red\color{black}).\\}
		\label{Fig:timeseriesFirst}
	\end{minipage}
	\vskip .1cm
	\begin{minipage}{\columnwidth}
		\centering
		\includegraphics[width=0.8\linewidth]{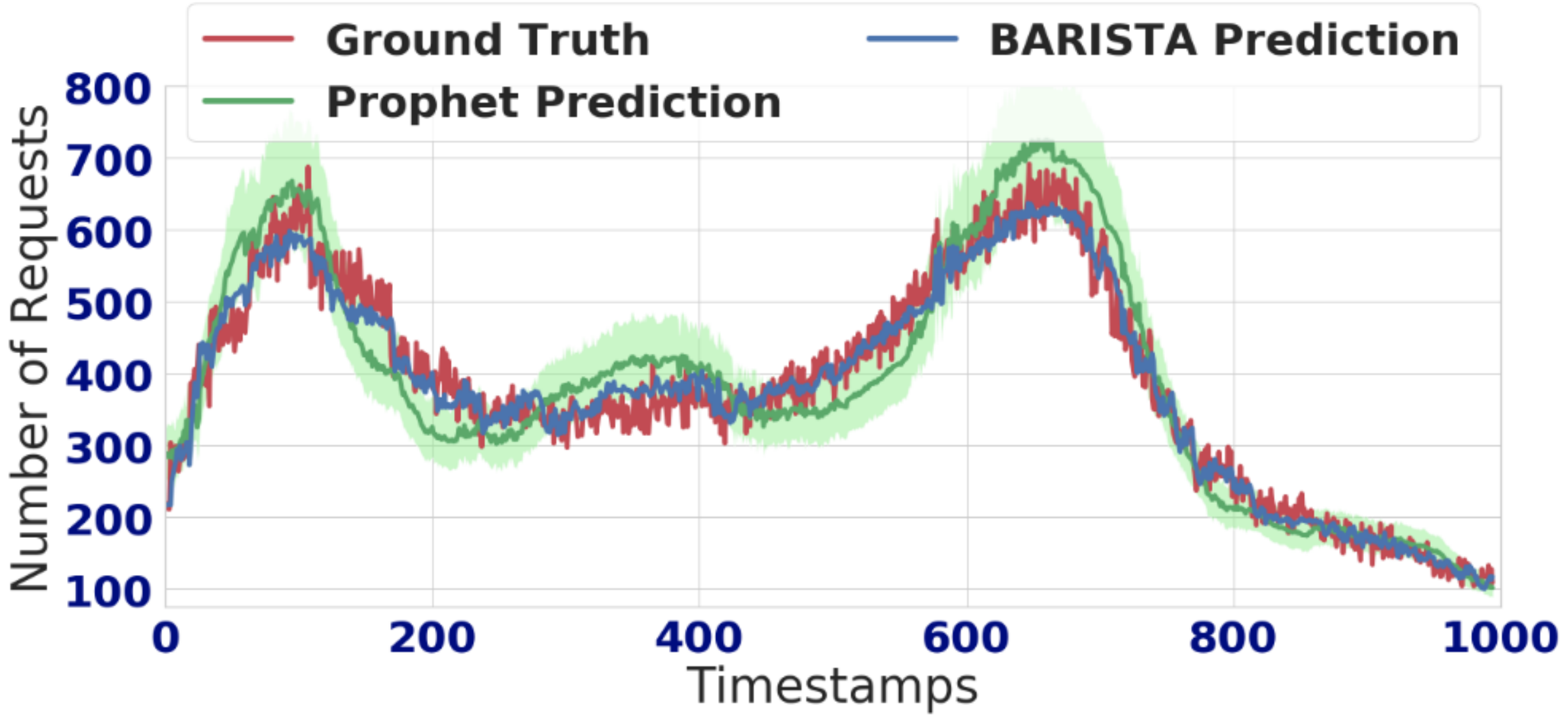} \vskip -2mm
		\captionsetup{width=0.95\linewidth}
		\caption{Performance comparison of Barista (\color{black!20!blue}blue\color{black}) and Prophet (\color{black!50!green}green\color{black}) along with ground truth (second dataset) (\color{black!30!red}red\color{black}).}
		\label{Fig:timeseriesSecond}
	\end{minipage}
\end{figure}

In Barista, we extended Prophet with a machine learning based compensator to adjust the forecast based on the last five prediction errors. We decided to use five errors based on empirical results. Apart from these five prediction errors, the recent forecast of Prophet along with the upper and lower estimation bounds are used as features for learning the model (see Equation~(\ref{eq:COMP})). We used 3000 data points of Prophet to train the compensator model, and another 1000 data points to test and analyze our hybrid approach. We used H2O's implementation of AutoML framework~\cite{automl} to identify the best family of learning algorithms and tune the hyper-parameters. We found that XGBoost-based gradient boosted trees outperformed other machine learning models such as neural networks, random forest model, and was selected as the best model. The mean absolute errors for training, cross-validation and testing in the first and second datasets are (12.65, 15.10, 21.26) and (12.24, 15.13, 22.65), respectively.

\begin{figure}[htb]
	\begin{minipage}{\columnwidth}
		\centering
		\includegraphics[width=0.85\linewidth]{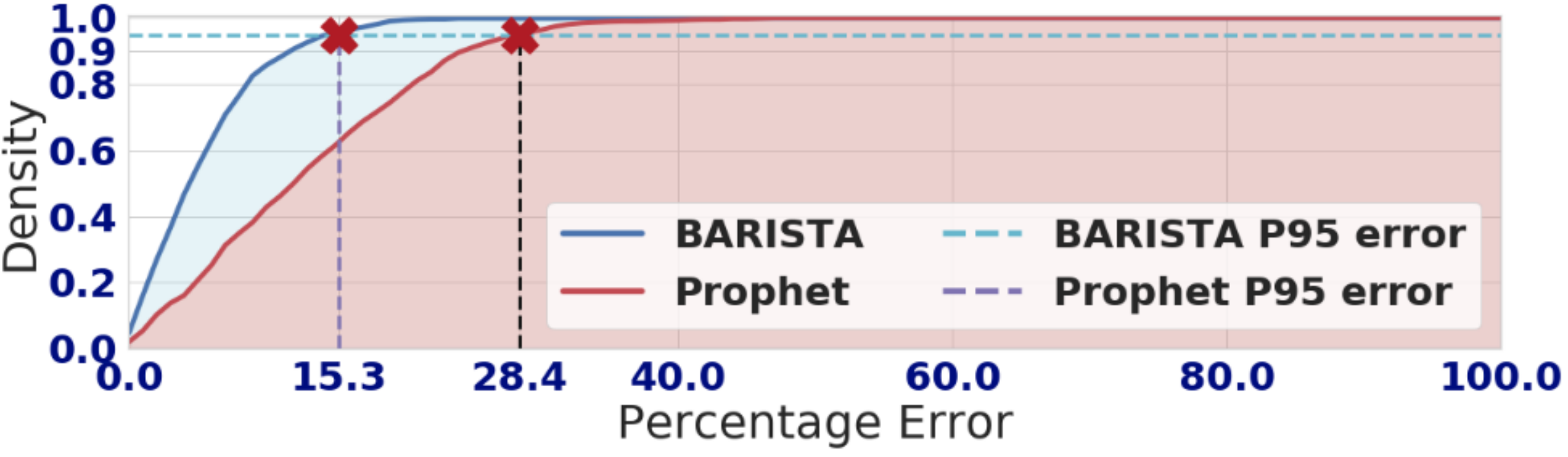}\vskip -2mm
		\captionsetup{width=0.95\linewidth}
		\caption{Cumulative absolute percentage error distribution of first dataset.\\}
		\label{Fig:cdfFirst}
	\end{minipage}
	\vskip .1cm
	\begin{minipage}{\columnwidth}
		\centering
		\includegraphics[width=0.85\linewidth]{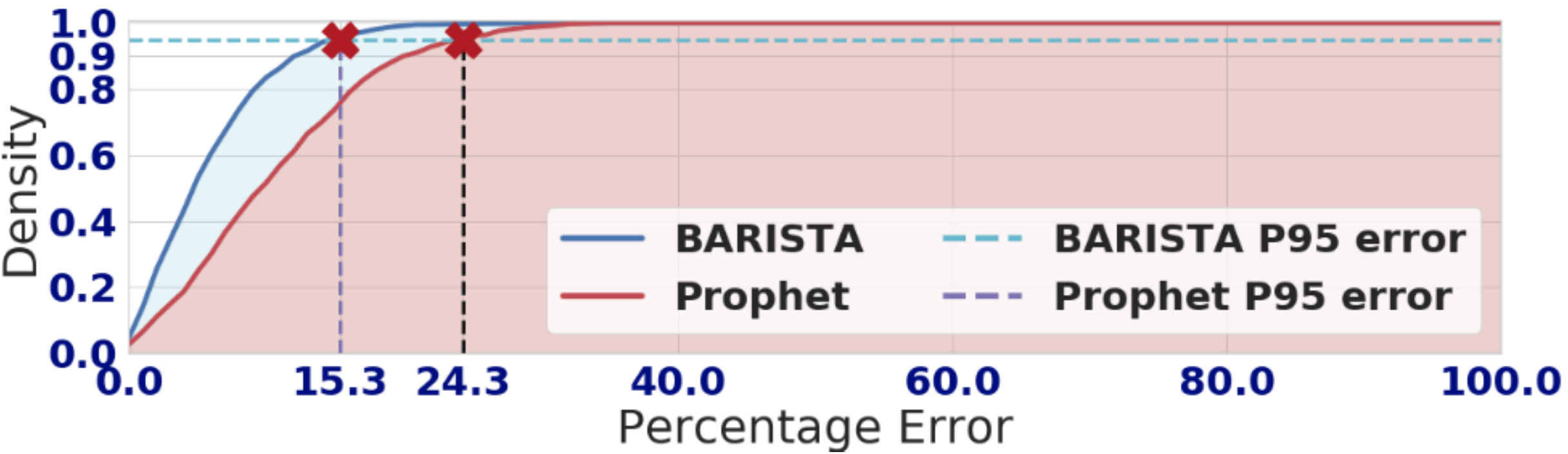}\vskip -2mm
		\captionsetup{width=0.95\linewidth}
		\caption{Cumulative absolute percentage error distribution of second dataset.}
		\label{Fig:cdfSecond}
	\end{minipage}
\end{figure}

The forecasting results of Barista and Prophet along with the actual workloads on the test dataset are shown in Figures~\ref{Fig:timeseriesFirst} and~\ref{Fig:timeseriesSecond}.
It is visible from the figures that the Barista prediction curves closely resemble the actual workloads and it predicts the sudden bursts of requests with more accuracy compared to Prophet which often lags and leads. Figures~\ref{Fig:cdfFirst} and~\ref{Fig:cdfSecond} show the cumulative percentage error distributions for both approaches. Barista outperforms Prophet by 37\% and 46\% in the first and second datasets, respectively.

%

%
%
%

\subsection{Resource Selection and Provision}

\begin{figure*}[tbh]
	
	\centering
	\begin{subfigure}[t]{0.32\textwidth}
		\centering
		\includegraphics[width = \linewidth]{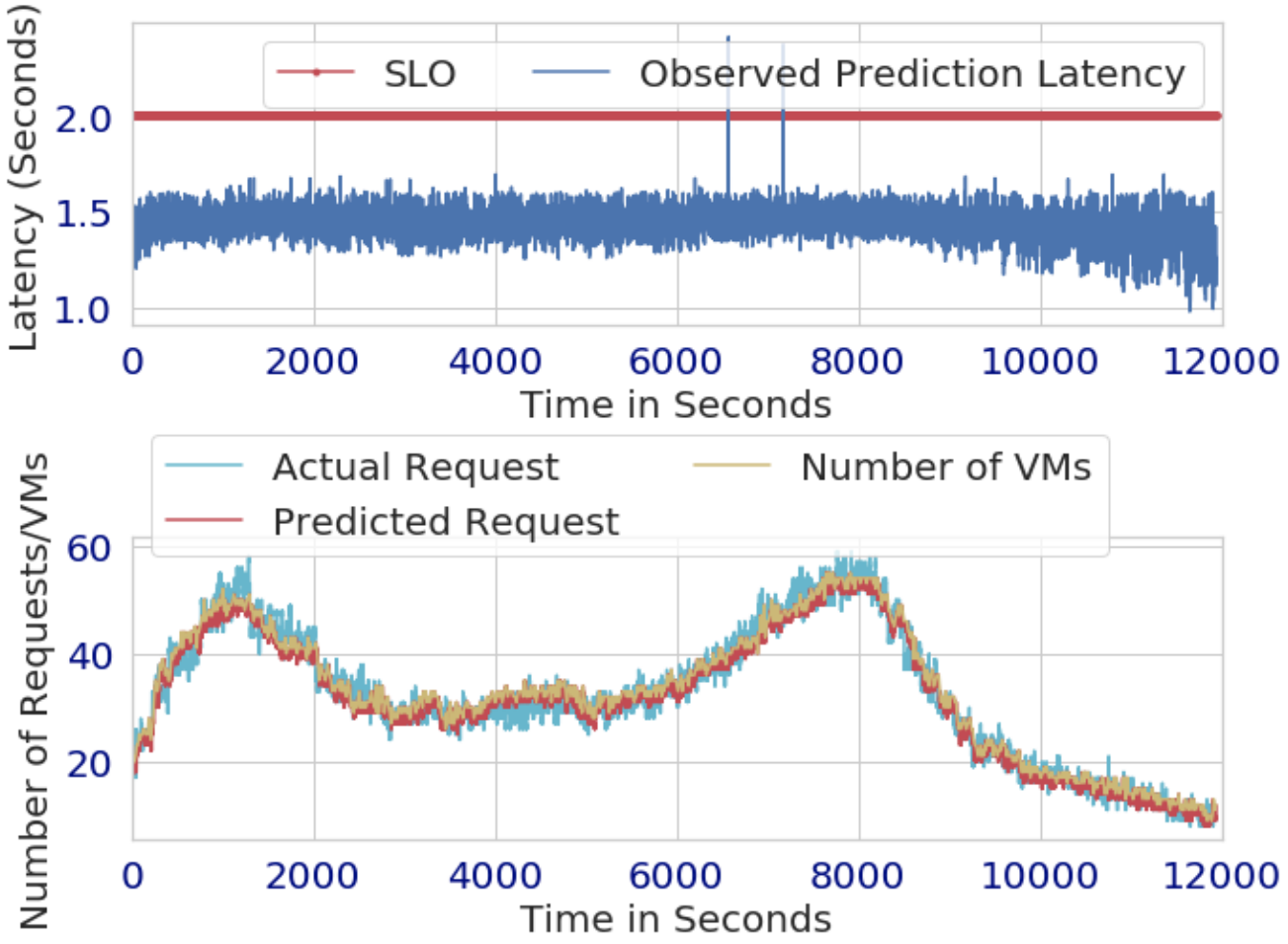}\vskip -1mm
		\caption{\scriptsize Upper image shows how we guaranteed the 2-second SLO for the Resnet prediction service and the experienced latency by using the VM configuration selected by Barista on toll dataset. Lower image shows the actual request rate, predicted request rate, and number of allocated VMs (t3.small (2cores)).}
		\label{Fig:resnet2}
	\end{subfigure}%
	~
	\centering
	\begin{subfigure}[t]{0.32\textwidth}
		\centering
		\includegraphics[width = \linewidth]{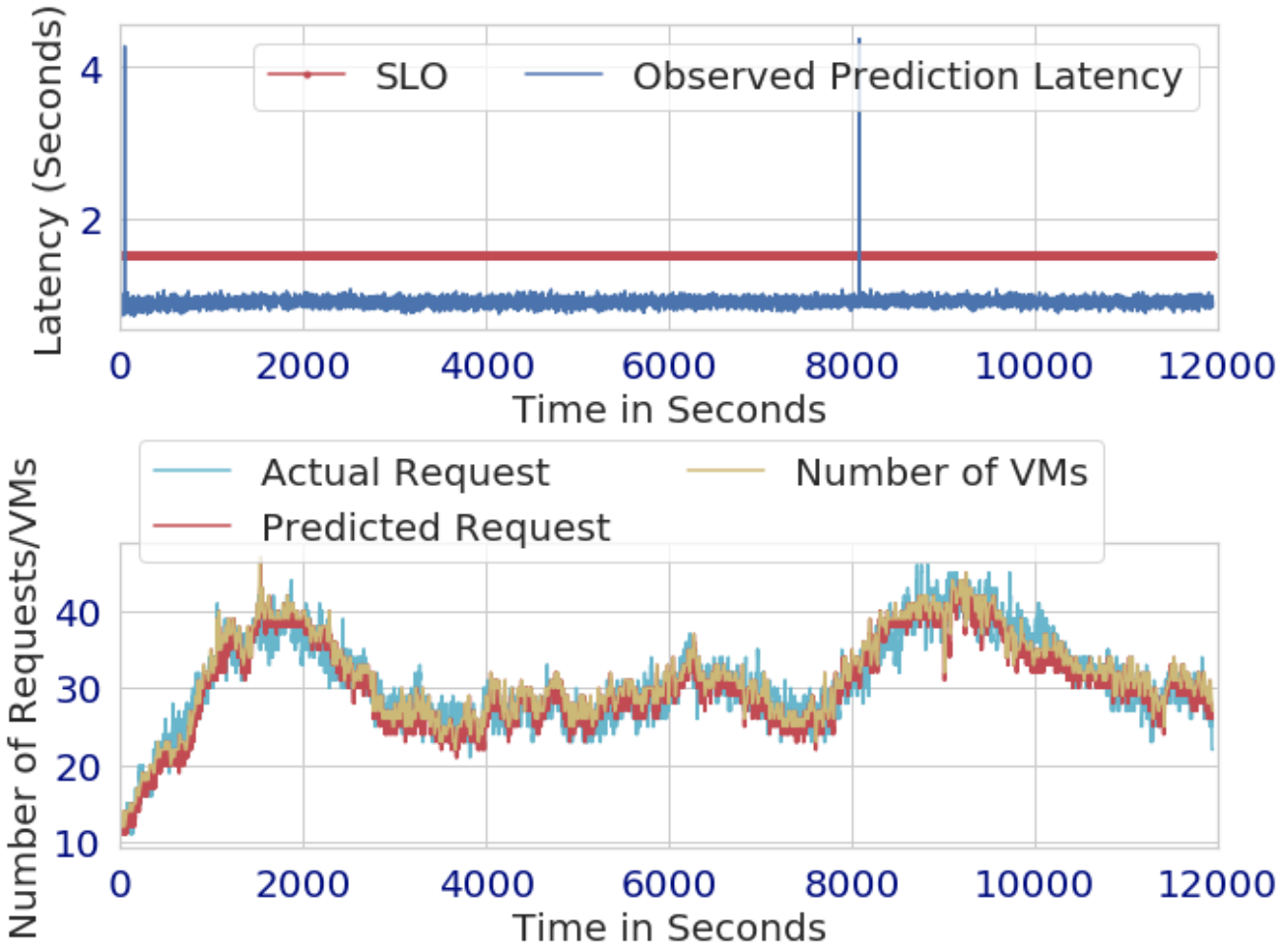}\vskip -1mm
		\caption{\scriptsize Upper image shows how we guaranteed the 1.5-second SLO for the Wavenet prediction service and the experienced latency by using the VM configuration selected by Barista on taxi dataset. Lower image shows the actual request rate, predicted request rate, and number of allocated VMs (t3.small (2cores)).}
		\label{Fig:wavenet15}
	\end{subfigure}%
	~
	\begin{subfigure}[t]{0.32\textwidth}
		\centering
		\includegraphics[width = \linewidth]{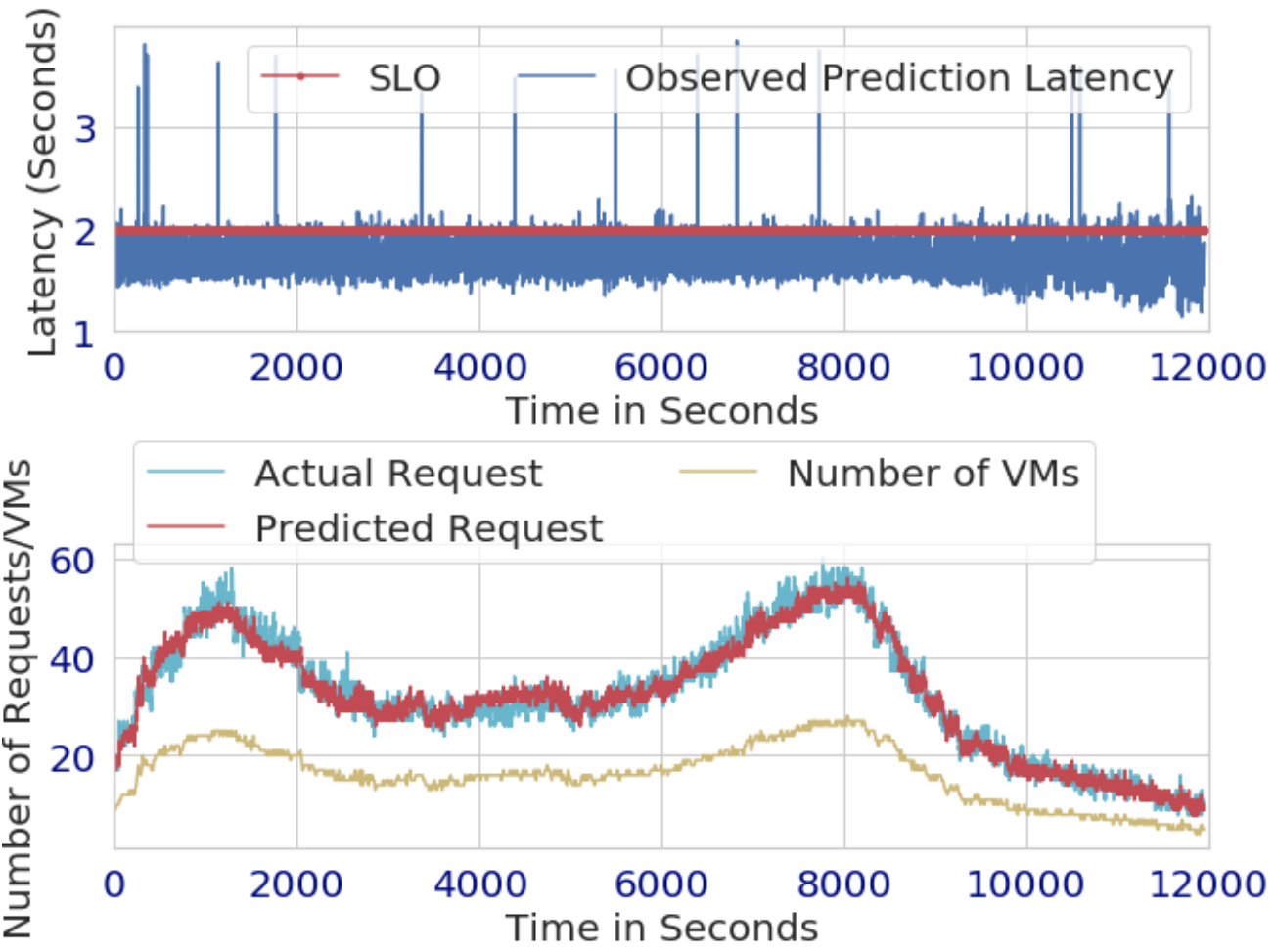}\vskip -1mm
		\caption{\scriptsize Upper image shows how we guaranteed the 2-second SLO for the Xception prediction service and the experienced latency by using the VM configuration selected by Barista on taxi dataset. Lower image shows the actual request rate, predicted request rate, and number of allocated VMs (t3.xlarge (4cores)).}
		\label{Fig:xception2}
	\end{subfigure}%
	\caption{Barista performance results on selected VM configuration as backend. }\vskip -5mm
	\label{Fig:Baritaperf}
\end{figure*}

Barista makes resource selection and provision decisions based on the algorithms described in Sections~\ref{sec:vm_deploy} and~\ref{provisioner}.  We evaluated our resource estimator and provisioner at different time points in the life cycle of the prediction services. We uniformly distributed the workload traces from one minute to five seconds in our experiment.  We met the target SLOs (2 seconds and 1.5 seconds) 99\% of the time over 12000 seconds for the Resnet and Wavenet services as shown in Figures~\ref{Fig:resnet2} and~\ref{Fig:wavenet15}. The SLO (2 seconds) compliance rate marginally dropped to 97$\%$ for the Xception service as shown in Figure~\ref{Fig:xception2}.


\begin{figure}[h]
	\centering
	\includegraphics[width=0.75\linewidth]{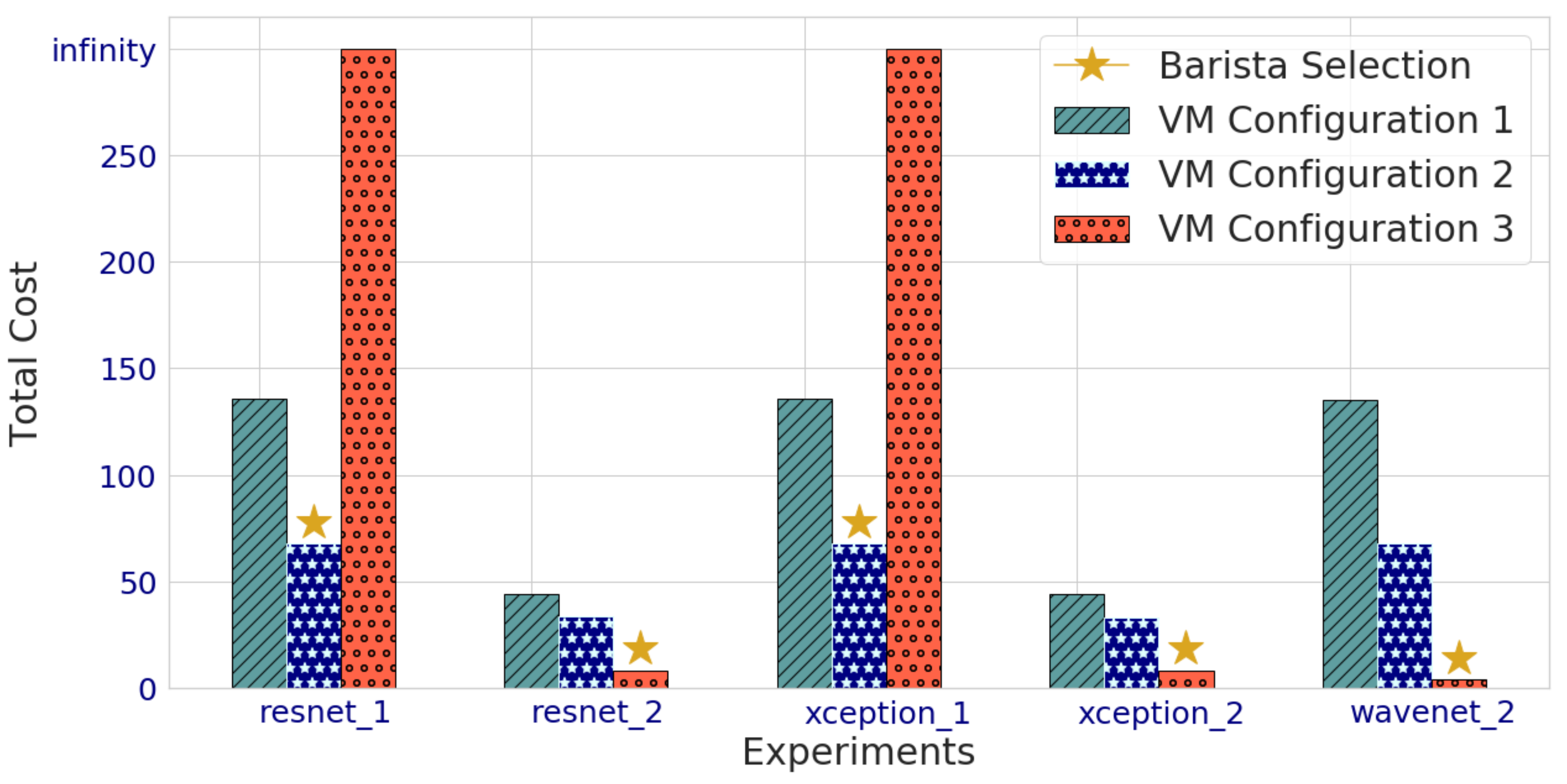}\vskip -1mm
	\caption{Cost comparison between multiple VM configurations (cost infinity means the VM is infeasible option; it cannot serve the request within the SLO bound).}
	\label{Fig:cost}
\end{figure}

Barista not only guarantees SLO compliance but also minimizes the running and management cost by intelligently selecting the VM configuration. 
As described in Section~\ref{sec:vm_deploy},
Barista selects the VM configuration depending on the SLO, the cost, and the estimated execution time of a prediction service. We emulated the prices of the VMs according to the Amazon EC2 instances and considered VM expiration time on an hourly basis (instance hour as an example scenario). Figure~\ref{Fig:cost} shows the total cost of hosting the backend VMs for 10 hours (600 mins) while guaranteeing the SLO bound for the workload traces~\cite{nycdata,nycthru}. Here, Configuration 1 represents t3.2xlarge, Configuration 2 represents t3.xlarge, and Configuration 3 represents t3.small (our $min\_mem$ constraint is 2GB). We solved the optimization problem for a given SLO bound to select a VM type, and considered this VM type as one of the configurations for our experiment. We then considered two other VM types of the same VM group (here, a group means Amazon EC2 instance type group, e.g., general purpose (t3, compute-intensive (c4-5 group)) with different core capacities.
Even if assigning more cores reduces the running time of a prediction service, because of the cost difference of different VM configurations, the naive approach of selecting the VM with the highest number of cores or the most powerful VM is not always the best option, as more VMs from a less powerful type can be more cost effective. This happens because the prediction services are stateless and utilize all the cores available in that VM, and requests are always served sequentially. We observe from Figure~\ref{Fig:cost} that Barista performs 50-95\% better than the naive approach.


\subsection{Reactive Vertical Scaling for Model Correction}

We monitored the latency of the services every five seconds and make decisions accordingly based on the monitored latency and the SLO bound. We considered a deployment scenario where prediction services can run with other co-located low-priority batch jobs. In this experiment, the goal is to demonstrate that we can vertically allocate and de-allocate CPU cores by monitoring the SLO.

\begin{figure}[tbh]
	\centering
	\includegraphics[width=\linewidth]{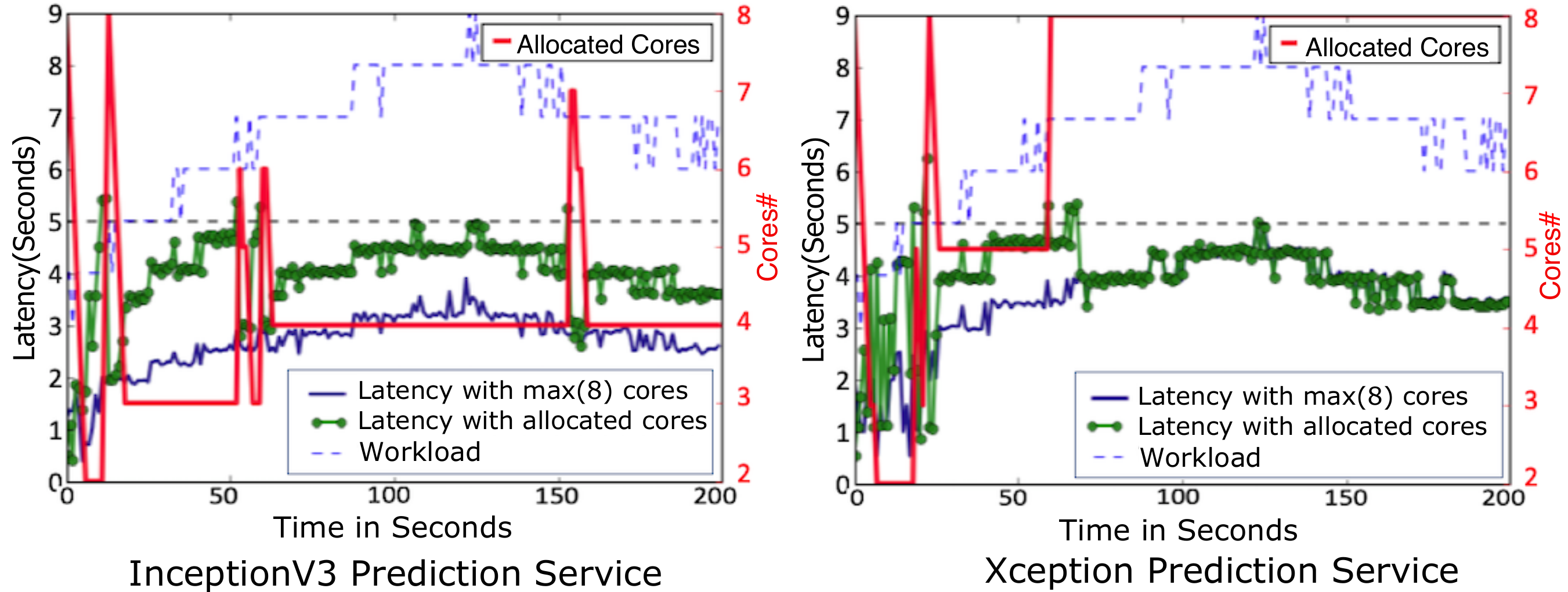}\vskip -2mm
	\caption{
		The The \color{blue}{blue} \color{black}  dotted line shows the workload pattern, and the solid \color{blue!50!black}{navy blue} \color{black}  line shows the latency of the prediction services 
		on a VM of 8 cores. The  \color{green!60!black}{green} \color{black} line shows the latency if we dynamically (de)-allocate the cores.}
	\label{Fig:vertical}
\end{figure}

Figure~\ref{Fig:vertical} shows how over-provisioning for two prediction services (Xception and
InceptionV3) could be handled reactively by adjusting the CPU cores at runtime. Using our reactive approach for vertical scaling, we saved approximately 15$\%$ and 30$\%$ of CPU shares of an eight cores OpenStack VM for Xception and InceptionV3, respectively, on a particular workload trace. Our reactive approach also achieves over 98\% of the SLO hits while optimizing the CPU shares significantly. We also observed similar behaviors for other prediction services, thus demonstrating the capabilities of Barista to make model correction if there is any resource over-provisioning due to over-estimated time-series prediction.



\section{CONCLUSION}
\label{sec:conclusion}


\textbf{\textit{Summary}}: 
Predictive analytics services based on deep learning pre-trained models can be hosted using serverless computing paradigm due to their stateless nature. However, meeting their service level objectives (SLOs), i.e., bounded response times and bounded hosting costs, is a hard problem because workloads on these services can fluctuate, and the state of infrastructure 
can result in different performance  characteristics. To resolve these challenges, this paper describes Barista, which is a dynamic resource management framework for providing horizontal and vertical autoscaling of containers based on predicted service workloads.

\textbf{\textit{Discussions}}: 
The Barista approach can be broadly applied to other compute-intensive and parallelizable simulation services, where the simulation model needs to be loaded in memory, and the simulation behavior is determined based on user requests and user-specified SLO. In this paper, 
we did not consider running different co-located prediction services together on the same machine, where workload patterns for each prediction services can be different.
Vertical scaling the different prediction services at the same time on the same machine is a dimension of our future work.




\section*{Acknowledgment}
\footnotesize
This work was supported in part by NSF US Ignite CNS 1531079, AFOSR DDDAS FA9550-18-1-0126 and AFRL/Lockheed Martin's StreamlinedML program. Any opinions, findings, and conclusions or recommendations expressed in this material are those of the author(s) and do not necessarily reflect the views of these funding agencies.

\bibliography{References}
\bibliographystyle{IEEEtran}

\end{document}